\documentclass[twocolumn,showpacs,preprintnumbers,amsmath,amssymb,prl]{revtex4-1}

\usepackage{dcolumn}
\usepackage{bm}

\usepackage{color}

\usepackage{graphicx}
\usepackage{natbib}
\usepackage{subfigure}
\usepackage{cases}
\usepackage{epstopdf}

\newcommand{\beq}{\begin{equation}}
\newcommand{\eeq}{\end{equation}}
\newcommand{\curl}{\nabla\times}

\begin{document}
\def\bfB{\mbox{\bf B}}
\def\bfQ{\mbox{\bf Q}}
\def\bfD{\mbox{\bf D}}
\def\etal{\mbox{\it et al}}
\def\bt{\tilde{\bf b} }
\title{Optimal lengthscale for a turbulent dynamo}
\author{Mira Sadek$^{1,2}$, Alexandros Alexakis$^{1}$, Stephan Fauve$^1$}

\affiliation{$^1$Laboratoire de Physique Statistique, Ecole Normale Sup\'erieure, CNRS, Universit\'e P. et M. Curie, Universit\'e Paris Diderot, Paris, France, 
             $^2$CRSI, Lebanese University, Hadath, Lebanon}

\date{\today}

\begin{abstract}
We demonstrate that there is an optimal forcing length scale for low Prandtl number dynamo flows,
that can significantly reduce the required energy injection rate. The investigation is based on simulations
of the induction equation in a periodic box of size $2\pi L$. 
The flows considered are turbulent ABC flows forced at different forcing wavenumbers $k_f$
simulated using a subgrid turbulent model. 
The critical magnetic Reynolds number $Rm_c^T$ decreases as the forcing wavenumber $k_f$ increases from the smallest allowed $k_{min}=1/L$.
At large $k_f$ on the other hand, $Rm_c^T$ increases with the forcing wavenumber as $Rm_c^T \propto \sqrt{ k_f}$ in agreement with mean-field scaling prediction. 
At $k_f L\simeq 4$ an optimal wavenumber is reached where $Rm_c^T$ obtains its minimum value.
At this optimal wavenumber $Rm_c^T$ is smaller by more than a factor of   
ten than the case forced in $k_f=1$. This leads to a reduction of the energy injection rate by three orders of magnitude 
when compared to the case that the system is forced in the largest scales and thus provides a new strategy for the design of a fully turbulent experimental dynamo.

\end{abstract} 

\maketitle

Dynamo is the mechanism by which magnetic fields are amplified in stars and planets
due to their stretching by the underlying turbulent flow \cite{Moffatt1978}. 
In the last two decades several experimental groups have attempted to reproduce the dynamo
instability in the laboratory \cite{Gailitis2001,Stieglitz2001,Vks2007}. The first successful dynamos were achieved 
in Riga \cite{Gailitis2001} and Karlsruhe \cite{Stieglitz2001}. The flow in these dynamos were highly constrained 
and did not allow for turbulence to fully develop at large scales. The first fully turbulent dynamo
was achieved in \cite{Vks2007} where the flow was driven by two counter rotating propellers. 
However, in this experiment dynamo was only obtained when at least one ferromagnetic iron propeller was used. 
So far other attempts to achieve dynamo 
are not successful and unconstrained dynamos
driven just by the turbulent flows have not been achieved.

One of the major difficulties to achieve liquid metal dynamos are the low values of magnetic 
Prandtl numbers $P_{_M}$ (the ratio of viscosity $\nu$ to magnetic diffusivity $\eta$) that characterizes liquid metals
and is less than $10^{-5}$. This implies that very large values of the Reynolds number
$Re=UL/\nu$ (where $U$ is the rms velocity and $L$ is the domain size) 
are needed to reach even order one values of the magnetic Reynolds numbers $Rm=UL/\eta=P_{_M} Re$.
The magnetic Reynolds number is the critical parameter that determines the dynamo onset.
For small values of $Rm$ no dynamo instability exists 
and $Rm$ should be larger than a critical value $Rm_c$ in order to generate the spontaneous growth of the magnetic field.
The large $Re$ needed to reach values of $Rm$ above $Rm_c$ imply large power consumption that scales like 
$I \propto \rho U^3L^3/\ell_f $
(where $\ell_f$ is the length scale of the forcing). 


The dependence of the dynamo threshold $Rm_c$ on the Prandtl number was investigated by 
different groups \cite{Ponty2005,Mininni2007,Scheko2005,Scheko2007} for different flows with the use of numerical simulations. 
%
%
These studies showed that as the magnetic Prandtl number is decreased the critical magnetic Reynolds number is 
initially increased. 
The turbulent fluctuations generated at large values of $Re$ 
inhibit dynamo action raising the critical power to values much larger than in the case of more organized laminar flows.
However when sufficiently large Reynolds numbers are reached this increase saturates  
and a finite value of $Rm_c$ is reached in the limit of $Re \to \infty$.
We will refer to this value as the turbulent critical magnetic Reynolds number 
and define it as $Rm_c^T\equiv \lim_{Re\to\infty}Rm_c$.  
The afore mentioned studies managed to reach this asymptote only by using subgrid scale models 
(either hyperviscocity, $\alpha$ model LES, or a dynamical turbulent viscosity) that model the high Reynold number flows. 
It is worth pointing out that the different flows considered led to different values of $Rm_c^T$
implying that it is possible to optimize the flow to reduce $Rm_c^T$.
This is what we investigate in this work by varying the length scale of 
the forcing $\ell_f$ with respect to the domain size $L$.  
A similar study but for laminar flows was performed in \cite{Tilgner1997,Plunian2002,Plunian2005}.
The study is based on the results of numerical simulations using a pseudospectral
method in a triple periodic domain \cite{Minini_code1,Minini_code2} and analytic estimates based on scale separation arguments.  
 
In their simplest form the governing equation for the evolution of the magnetic field is given by
\beq
\partial_t {\bf b} = \curl ({\bf u \times b} ) + \eta \Delta {\bf b}
\label{ind}
\eeq
where ${\bf b}$ is the magnetic field,  and $\eta$ the magnetic diffusivity. 
${\bf u}$ is the velocity field that is determined  by solving 
the independent incompressible Navier-Stokes equation of a unit density $\rho=1$ fluid,  
\beq
\partial_t {\bf u} + ({\bf u \cdot \nabla u} ) = -\nabla P + \nu \Delta {\bf u} +{\bf f}.
\label{NS}
\eeq
where ${\bf f}$ is an external forcing. In the present study the domain considered
is a triple periodic box of size $2\pi L$ and $\bf f$ is taken to be the ABC forcing 
\[
   {\bf f}= \left[ \begin{array}{c}
          A \sin(k_f z)+C \cos(k_f y), \\ 
          B \sin(k_f x)+A \cos(k_f z), \\ 
          C \sin(k_f y)+B \cos(k_f x )
            \end{array} 
            \right] \]  with $A=B=C=1$.
$k_f$ is the forcing wave number, and we define the forcing lengthscale as $\ell_f \equiv k_f^{-1}$.  
The ABC flow has been the subject of many dynamo studies both in the laminar \cite{Childress1970,Alexakis2011} and turbulent state \cite{Mininni2007}. 

In this set up some analytical progress can be made in the case that $k_f L\gg 1$. 
Then one can use standard mean-field approximations  to estimate the critical onset $Rm_c^T$ 
\cite{Parker1955,Steenbeck1966,Childress1970}.
Splitting the magnetic field in a large scale component $\bf B$ and a fluctuating part $\bt$
for scale separation we obtain to first order for the fluctuating field
\beq
\eta \Delta \bt = -{\bf B \nabla u} 
\label{sse}
\eeq  
and for the large scale field
\beq
\partial_t {\bf B} = \nabla \times (\alpha {\bf B} ) - \eta \Delta {\bf B}
\eeq
where $\alpha$ is a tensor such that $\alpha_{ij}  B_j = \langle {\bf u \times \bt} \rangle_i$
the angular brackets stand for small scale average and summation over the index $j$ is implied. 
For the particular forcing chosen
the $\alpha$-tensor is diagonal and isotropic.
The non-zero diagonal elements $\alpha_{ii}$ can be calculated by solving (\ref{sse}) for a given velocity field and are
\beq
\alpha_{ii} = \frac{-1}{\eta} \langle {\bf u} \times \Delta^{-1} \nabla_i {\bf u} \rangle_i = a \frac{U^2 }{\eta k_f}, 
\eeq
(no summation over $i$ is implied).
$a$ is the non-dimensional $\alpha_{ii}$ element that is independent of $U,\eta$ and $k_f$
and needs to be calculated from the flow.  
The growth rate $\gamma $ for a helical large-scale mode of wavenumber $K=1/L$ is then
\beq
\gamma = a \frac{U^2 }{\eta k_f L} - \eta L^{-2} 
\eeq 
that leads to a critical Reynolds number
$Rm_c = \sqrt{k_fL/a}$.
This result was discussed in \cite{Fauve2002}.
Note that this argument is true both for turbulent and laminar flows 
although the value of the coefficient $a$ will depend on the level of turbulence. 
At large $Re$ however $a$ will reach an asymptotic value $a^{_T}$ that will determine 
the value of $Rm_c^T$ to be 
\beq 
Rm_c^T=\sqrt{k_fL/a^{_T}}. \label{sqr}
\eeq
This scaling implies that in the large scale separation limit $k_fL\gg1$
for fixed $L$ and $\eta$ as $k_f$ is increased it becomes more difficult to obtain   
a dynamo. 
 
To calculate $Rm_c^T$ in the absence of scale separation 
we performed numerical simulations varying the forcing wavenumber $k_f$ for fixed $L$
of equations \ref{ind},\ref{NS}.
In order to mimic the large Reynolds number flow that requires large grid size $N$
we do not use an ordinary viscosity $\nu$ for the dissipation but rather
a dynamical wavenumber-dependent turbulent viscosity \cite{Chollet1981} defined 
in spectral space as
\beq
\nu_{_T}(k,t)    = 0.27[1+3.58(k/k_c)^8]\sqrt{E_{_K}(k_c,t)/k_c}
\eeq
where $k=|{\bf k}|$ is the wavenumber, $k_c=N/3L$ is the maximum wavenumber after de-aliasing and $E_{_K}(k,t)$ is the kinetic energy  
spectrum of the flow. The same modeling was also used in \cite{Ponty2005} to obtain $Rm_c^T$. 
$\nu_{_T}$ depends on the grid size $N$, but the dependence of the large scale 
components of the flow on $N$ are expected to die off much faster than in the case of ordinary viscosity.

To calculate $Rm_c^T$ as a function of $k_f$
the following procedure was used (see also \cite{Ponty2005,Scheko2007}): 
For a given $k_f$ and grid size $N$ a series of simulations were performed varying $Rm$
and the exponential growth rate of the magnetic energy was measured. 
The onset $Rm_c$ was determined by linearly interpolating the values of $Rm$ between the slowest growing dynamo
and the slowest decaying dynamo. The  series of runs was then repeated for higher values 
of $N$ until either the value of $Rm_c$ remained unchanged in which case this determined $Rm_c^T$ or
we reached the maximum of our attainable resolution. 
Typically convergence was reached at grid sizes $N=256$ but a few runs at $N=512$ were also performed 
for verification.
In addition we solved equation \ref{sse} (with $\bf u$ given by equation \ref{NS}) with an imposed uniform magnetic field to
calculate the elements of the $\alpha$ tensor and determine $a^{_T}$. 
\begin{figure}                                                                           %
\includegraphics[width=\linewidth]{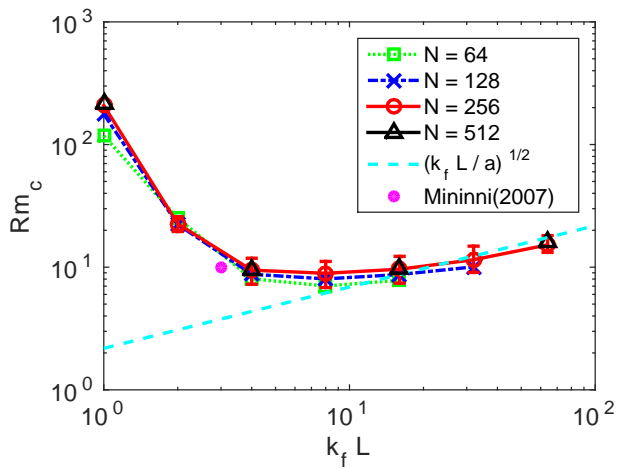}                                            %
\caption{\label{fig1} $Rm_c$ as a function of $k_fL$ obtained from numerical simulations %
of different resolutions $N=64-512$. The data indicate that for $N\ge128$, $Rm_c$        %
does not vary as $N$ is increased further and thus it approximates well $Rm_c^{_T}$.     %
The error-bars correspond to the maximum/minimum value of $Rm$ for which we obtained     %
a clear positive/negative growth rate for the simulations with $N=256$.                  %
The filled circle corresponds to the value of $Rm_c^{_T}$                                %
obtained from direct numerical simulations in \cite{Mininni2007}.                        %
The dashed line shows the                                                                %
mean field (alpha) prediction valid in the limit $kL\to\infty$.                  }       %
\end{figure}                                                                             %

The results for the critical 
Reynolds number as a function of the forcing wave number $k_f$ are shown in figure \ref{fig1}.
The predicted
scaling behavior (\ref{sqr}) is shown with a dashed line in the same figure \ref{fig1}.
The black dot indicates the the value of $Rm_c^T$  calculated in \cite{Mininni2007} using 
simulation of higher resolutions and $\alpha$-model LES.

The results are very motivating for future laboratory experiments. 
Although for large $k_fL$ the asymptotic scaling of (\ref{sqr}) seems to be verified 
indicating that making the forcing length scale very small will not benefit dynamo experiments,
at intermediate length scales $Rm_c^T$ appears to reach a minimum around $k_f L=4$ to $8$. 
In fact the value of $Rm_c^T$ at this optimal wavenumber is one order of magnitude smaller
than the value of $Rm_c^T$ at $k_fL=1$. 

\begin{figure}                                                                          %
\includegraphics[width=0.8\linewidth]{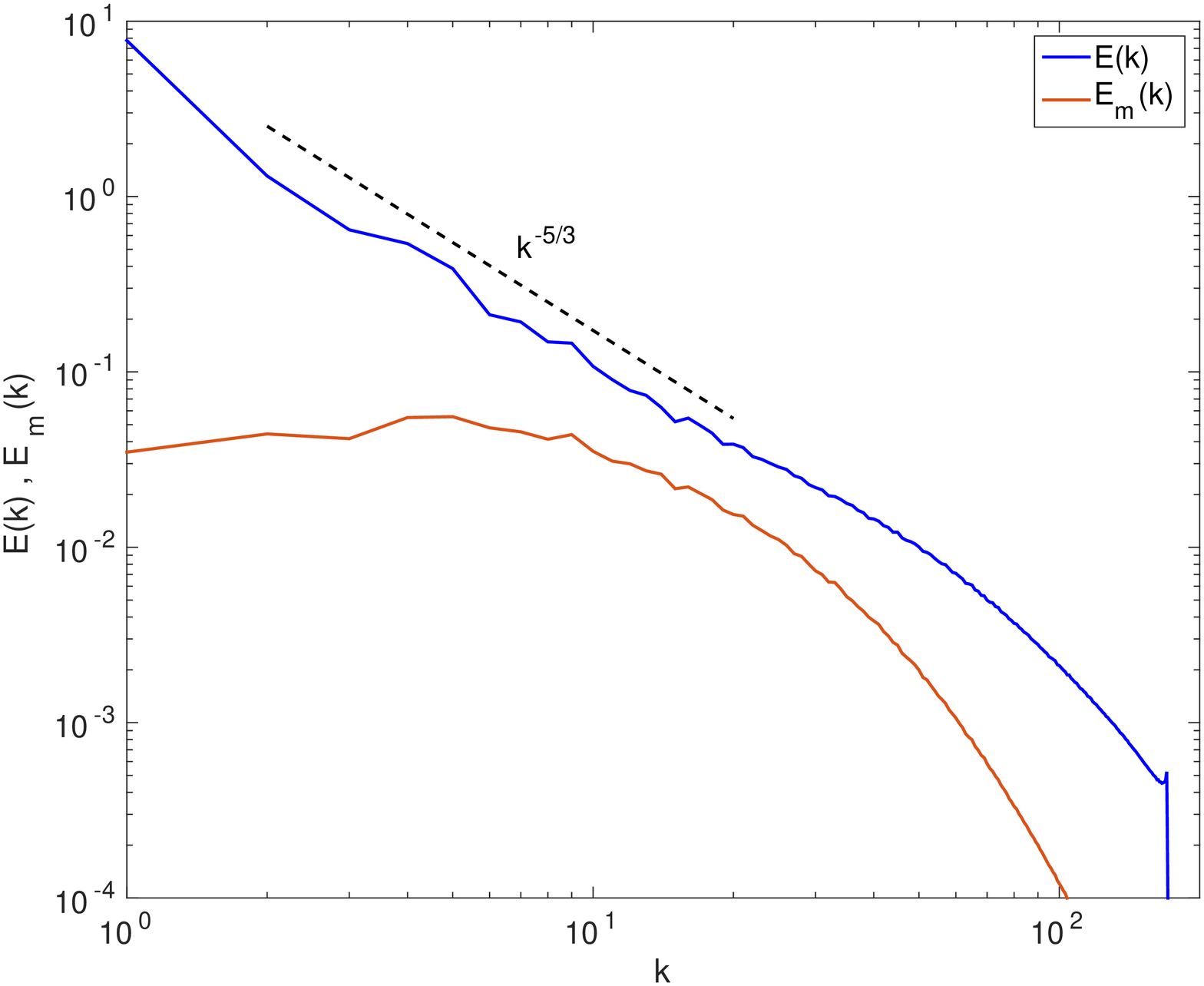}                                        %
\includegraphics[width=0.8\linewidth]{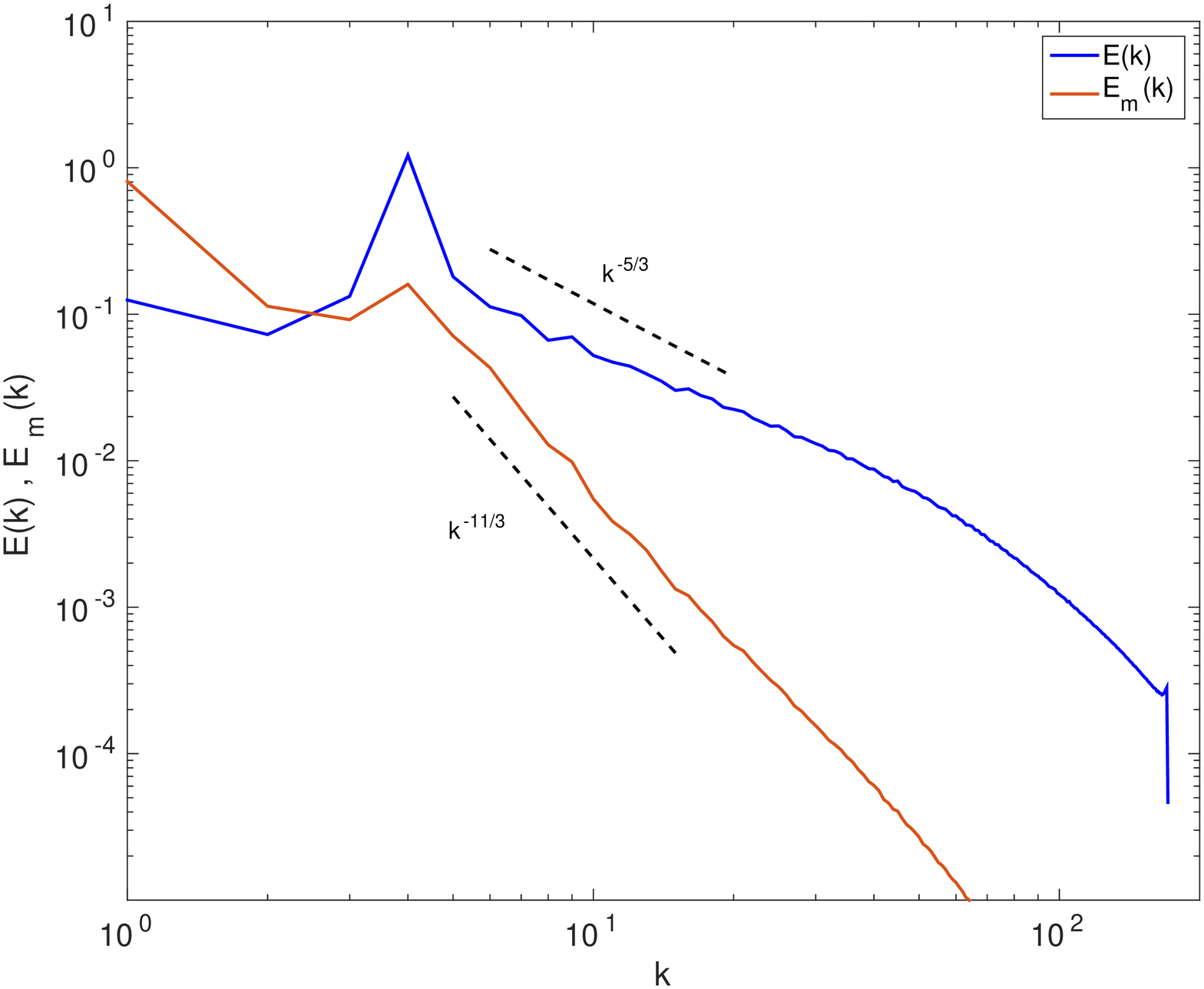}                                        %
\includegraphics[width=0.8\linewidth]{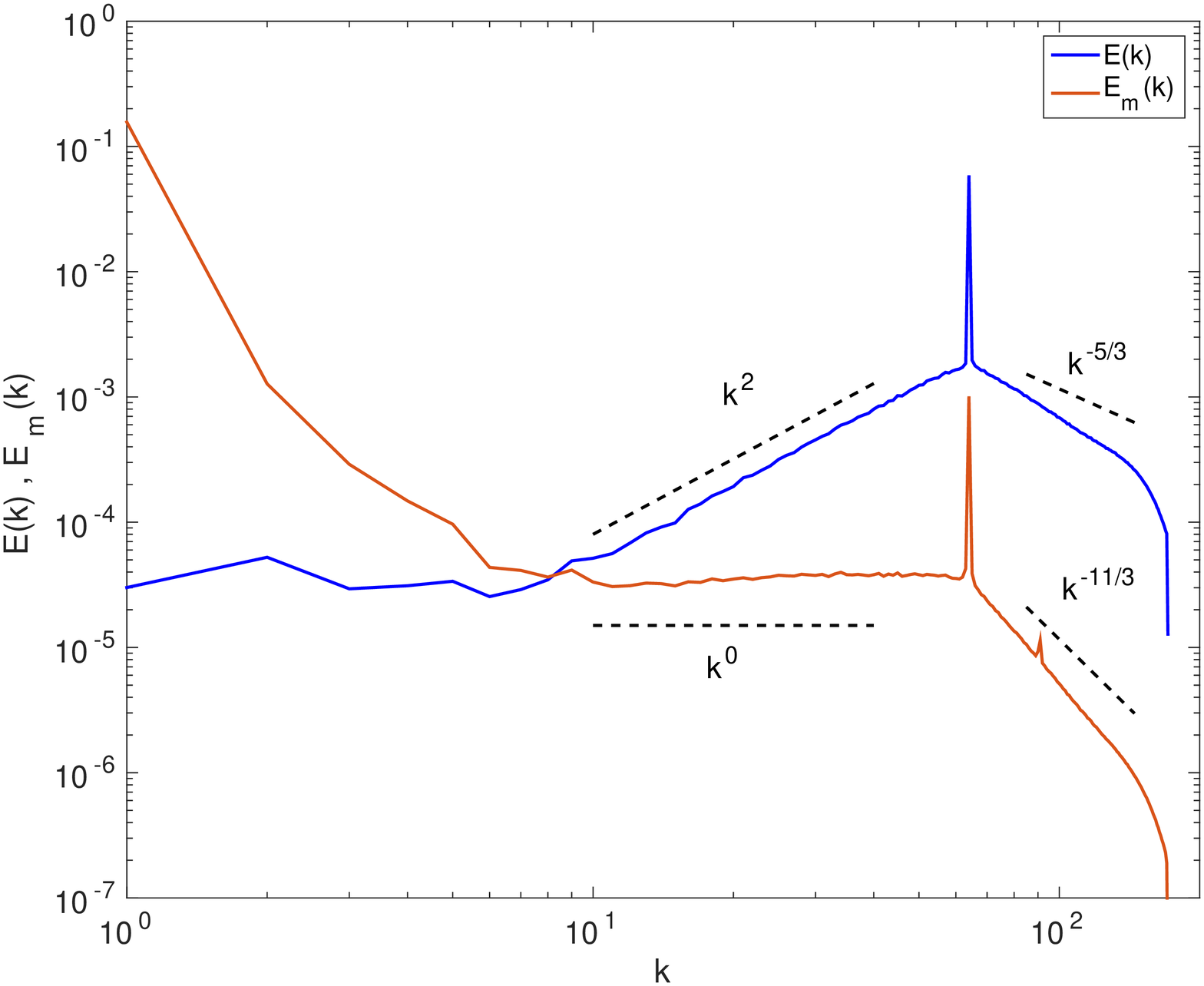}                                        %
\caption{\label{fig2} Kinetic $E(k)$ and magnetic $E_m(k)$ energy spectra               %
for the marginally unstable modes for $k_fL=1 $ (top),                                  %
                                      $k_fL=4 $ (middle),                               %
                                      $k_fL=64$ (bottom).                               %
}                                                                                       %
\end{figure}                                                                            %

The kinetic and magnetic energy spectra for slowest growing mode for
three different forcing wave-numbers at highest resolution $N=512$ are shown 
in the three panels of figure \ref{fig2}. 
For the case that the flow is forced in the largest scale $k_f L=1$  
the kinetic energy shows a clear $k^{-5/3}$ spectrum while no clear
power law scaling can be observed for the magnetic field.
Most of the magnetic energy is concentrated in the 
small scales with very weak energy in the largest scale. 

At the other extreme where the flow is forced in the small scales
$k_fL=64$ most of the kinetic energy is in the small scales 
with a $k^{-5/3}$ scaling in the sub-forcing scales and a $k^2$ 
power-law scaling for the scales larger than the forcing scale,
suggesting equipartition of energy among all modes as equilibrium statistical mechanics predict \cite{Kraichnan1973}. 
In addition a small peak at large scales $kL\simeq1-2$ is observed. 
The $k^2$ energy spectrum has been observed before in numerical simulations of the truncated Euler equations \cite{Brachet2009},
and more recently in simulations  forced in the small scale where the excess of energy in the largest scale of the system
has also been observed \cite{Gopal2014,Dallas2015}. The role and cause of this peak and its effect on dynamo is the subject of current
investigations. The magnetic field on the other hand has a dominant peak at $kL=1$, 
caused by the $\alpha$ dynamo that is followed by a flat spectrum $k^0$, a peak at the forcing scale 
and then by a $k^{-11/3}$ power law until the dissipation scales. 
The two peaks at $kL=1$ and $k=k_f$ are in agreement with the mean field dynamo prediction.  
The two power-laws can also be explained by a balance between the stretching rate $S$ of the large scale field ${B_{_L}}$
by the fluctuations $u_\ell$ that is proportional to $S\propto {B_{_L}}u_\ell/\ell$ 
and the Ohmic dissipation that is proportional to $\eta b_\ell /\ell^2$. 
Substituting $u_\ell \propto \ell^{1/3}$ for the turbulent scales and $u_\ell \propto \ell^{-3/2}$ for the scales in equipartition
one recovers the two exponents $k^{-11/3}$ and $k^0$ respectively. 
The $k^{-11/3}$ spectrum was  predicted in \cite{Golitsyn1960,Moffatt1961} and has been
observed in experiments \cite{Fauve1998} and numerical simulations \cite{Ponty2004}.
%
The flat spectrum $k^0$ up to our knowledge is reported for the first time here. 

The case $k_f L=4$ that is close to the optimal wavenumber seems to be somewhere is between the two extreme cases. 
The magnetic field  at the largest scale appears neither dominant as in the mean field case bu nor negligible as in the $k_fL=1$ case.
In the large scales there is not enough scale separation to observe any power-law, 
but in the small scales a power law close to $k^{-5/3}$  
for the kinetic spectrum and $k^{-11/3}$ for the magnetic spectrum can be seen.
%
\begin{figure}                                                                           %
\includegraphics[width=\linewidth]{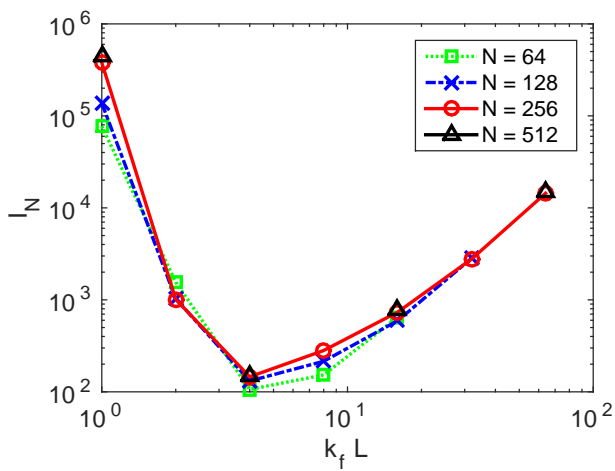}                                            %
\caption{\label{fig3} The non-dimensional critical injection rate $I_N$ as a function of %
 $k_fL$. }                                                                               %
\end{figure}                                                                             %

Our results have shown that future dynamo experiments can benefit from forcing at scales
smaller than the domain size by a factor of 4 to 8. To further demonstrate this fact     
in figure \ref{fig3} we plot the minimum energy injection rate 
$I =(2\pi L)^3 \langle {\bf f\cdot u} \rangle$
to achieve dynamo, normalized by the domain size $L$, the mass density $\rho$ and magnetic diffusivity $\eta$
$I_N=I L/(\rho \eta^3)$.
The reason we have chosen this non-dimensionalization is because the mass density  and magnetic diffusivity
are properties of the liquid metals that vary only with temperature, while the domain size is typically fixed.
In other words by normalizing it this way we ask the question: 
{\it in a laboratory experiment of a given the domain size what is the optimal forcing scale to achieve dynamo
 with a minimal the energy injection rate?}.     

The result is very encouraging! The optimal injection rate is almost three orders of magnitude
smaller than the case for which the forcing was in the largest scale. This large 
drop in the injection rate can be partly explained by considering the turbulent scaling 
for the energy injection rate $I \propto \rho U^3 L^3/\ell$. Substituting $U$ from the definition of
$Rm$ we obtain $I \propto \rho Rm^3 \eta^3/\ell$ and thus $I_N \propto Rm^3 k_fL$.
Thus the energy injection rate is very sensitive on changes in $Rm$ and the beneficial factor 
of 20 that was observed in figure \ref{fig1} translates to a factor 2000 for the energy injection rate.    
It is also worth pointing out that the while for optimizing $Rm_c^{_T}$ the optimal forcing wavenumber was 
between $k_fL=4$ and $k_fL=8$, in the case that $I_N$ is optimized the optimal wave number is more clearly the $k_fL=4$. 

In the light of this result we can envision the design of new dynamo
experiments where the flow is forced by an array of propellers so as 
to result in the small scale forcing required.
Such experiments will challenge long standing theoretical assumptions
of mean field dynamo theories.
 
\begin{acknowledgments}
This work was granted access to the HPC resources of GENCI-CINES (Project
No.x2014056421, x2015056421)
and MesoPSL financed by the Region Ile de France and the project Equip\@Meso
(reference ANR-10-EQPX-29-01). 
\end{acknowledgments}

\bibliographystyle{apsrev4-1}
\bibliography{dynamo}
\end{document}